\documentclass[prb,twocolumn,showpacs,aps,superscriptaddress,floatfix]{revtex4-2}
\usepackage{amsmath}
\usepackage{amssymb}
\usepackage{amsfonts}
\usepackage{bm}
\usepackage[dvips]{graphicx}
\usepackage{color}
\usepackage{ulem}
\usepackage[unicode, pdftex]{hyperref}
\usepackage{verbatim}
\begin{document}
\title{Kitaev-Ising-$J_1$-$J_2$ model: a density matrix renormalization group study}

\author{A.V. Kapranov}
\email[]{kapranov.av@phystech.edu}

\affiliation{Moscow Institute of Physics and Technology, Dolgoprudny,
    Moscow Region, 141701 Russia}

\author{R.S. Akzyanov}
\affiliation{Institute for Theoretical and Applied Electrodynamics, Russian
    Academy of Sciences, Moscow, 125412 Russia}

\begin{abstract}
We numerically study the Kitaev honeycomb model with the additional XX Ising interaction between the nearest and the next nearest neighbors (Kitaev-Ising-$J_1$-$J_2$ model), by using the density matrix renormalization group (DMRG) method. Such additional interaction correspond to the nearest and diagonal interactions on the square lattice. Phase diagram of the bare Kitaev model consist of low entangled commensurate magnetic phases and entangled Kitaev spin liquid. Anisotropic Ising interaction allows the entangled quantum paramagnetic phases in the phase diagram, which in the absence of the magnetic field previously was predicted for more complex type of interaction. We study the scaling law of the entanglement entropy and the bond dimension of the matrix product state with the size of the system. In addition, we propose an optimization algorithm to prevent DMRG from getting stuck in the low-entangled phases.

\end{abstract}
\maketitle

\section {Introduction}
Quantum spin liquids (QSLs) are strongly correlated phases of matter characterized by the absence of long-range magnetic order due to strong quantum fluctuations. As a result, electron spins do not form an ordered pattern but remain liquid-like even at zero temperature~\cite{Balents2010,Savary2016,Zhou2010,Wen2019,Broholm2020}. QSLs have been extensively studied due to their rich physics and unusual properties, including intriguing topological characteristics~\cite{Wen1991,Wen2013} and supporting fractional excitations~\cite{Senthil2000,Senthil2001}. QSLs play an important role in understanding strongly correlated materials~\cite{Read1991,Sachdev1992,Lee2006} and have the potential to serve as platforms for topological quantum computation in the future~\cite{Kitaev2006}.

The Kitaev honeycomb model~\cite{Kitaev2006} is an exactly solvable model whose phase diagram hosts both gapped and gapless $\mathbb{Z}_2$ quantum spin liquids. The exact solution demonstrates that, in the Kitaev spin liquid (KSL), spins fractionalize into emergent quasiparticles—Majorana fermions. One of the key advantages of the Kitaev model is its conceptual simplicity, which makes it a promising candidate for experimental realization. However, identifying material realizations of KSLs remains challenging due to the absence of a conventional order parameter~\cite{Zhou2017,Wen2019}. Recently, various iridate and ruthenate compounds have been proposed as potential platforms for realizing KSLs~\cite{Jackeli2009,Plumb2014,Yokoi2021,Banerjee2022,Bruin2022,Trebst2022,Nica2023}.

To explain Kitaev physics in real materials, numerous extended Kitaev models have been proposed and extensively studied. Historically, the first extension of the Kitaev model was the inclusion of the isotropic nearest-neighbor (NN) Heisenberg interaction $J$, which arises from the direct overlap of $d$-orbitals (Kitaev-Heisenberg model)~\cite{Jackeli2009,Chaloupka2010,Chaloupka2013,Yamaji2016,Gotfryd2017}. Later, it was shown in~\cite{Rau2014} that an additional bond-dependent anisotropic spin-exchange interaction exists between NN sites, commonly referred to as the $\Gamma$ coupling (Kitaev-Heisenberg-Gamma model). The $\Gamma$ interaction is highly frustrated and is the second strongest interaction in Kitaev materials, leading to a complex interplay with Kitaev interactions~\cite{Rau2014,Chaloupka2015,Lee2015,Yadav2016,Nishimoto2016,Yadav2019,Zhang2021,Buessen2021}. A more realistic model must also account for the effects of trigonal splitting of the $t_{2g}$ orbitals, which induces an additional anisotropic coupling $\Gamma'$, which is not symmetry-allowed in the absence of the trigonal distortion~\cite{Rau2014}. Other perturbations present in real materials have also been investigated, including the second-neighbor Kitaev interaction $K_2$~\cite{Rousochatzakis2015}, and the second- and third-neighbor Heisenberg interactions $J_2$, $J_3$~\cite{Kimchi2011,Sizyuk2014}.

Another approach to realizing exotic phases of condensed matter is through quantum simulations, where quantum-mechanical devices are used to mimic and investigate quantum matter~\cite{Buluta2009,Altman2021}. Quantum simulation provides a powerful framework for studying highly entangled phases, such as QSLs, where classical numerical methods fail. QSLs have been realized using Rydberg atom quantum simulators~\cite{Semeghini2021} (toric code-type QSLs), superconducting quantum processors~\cite{Satzinger2021} (Kitaev toric code), optical cavities~\cite{Chiocchetta2021}, and programmable quantum devices such as the D-Wave DW-2000Q~\cite{Zhou2021}. Recently, the Kitaev model on a square-octagon lattice was simulated using the variational quantum eigensolver on a superconducting processor~\cite{Li2023_1}. Another approach to studying Kitaev spin liquids was demonstrated in a recent work~\cite{Umeano2025}, where a quantum subspace expansion was performed, mapping a spin Hamiltonian onto a quantum simulator with the same coupling topology, effectively creating a quantum digital twin.

Despite significant progress in quantum computing and simulation, relatively high error rates in two-qubit gates remain a challenge~\cite{Krantz2019}. One of the major sources of gate errors is non-ideal interactions that induce parasitic coupling between nearest-neighbor (NN) and next-nearest-neighbor (NNN) qubits. These interactions are described by an XX (or YY) Ising-like interaction between NN and NNN qubits~\cite{Yan2018,Sung2021}. In this work, we map the Kitaev model onto a superconducting quantum processor with the same coupling topology as the Kitaev model. However, due to parasitic interactions between qubits, the resulting system deviates from the pure Kitaev model, forming an extended Kitaev model, which we analyze in detail.

The DMRG is a powerful numerical method for computing properties of one-dimensional (1D) quantum systems and is widely regarded as the most effective technique for 1D systems~\cite{White1992,White1993}. The DMRG algorithm optimizes a matrix product state (MPS)\cite{Affleck1987,Ostlund1995,Rommer1997}, a robust ansatz represented as a 1D tensor network. The computational cost of DMRG for a system with $N$ sites scales as $O(ND^3)$, where $D$ is the bond dimension, which determines the size of the MPS and its capacity to capture quantum entanglement. The key reason for the success of DMRG is that the ground states of local Hamiltonians obey the area law~\cite{Srednicki1993,Vidal2003,Gioev2006,Amico2008}. This implies that the ground states of local Hamiltonians can be accurately represented by MPS with a fixed bond dimension $D$, which grows exponentially with the entanglement entropy $S$, where $S$ scales with the system size $N$. Consequently, the computational cost is constrained by the area law of entanglement entropy. Thus, DMRG is highly efficient for 1D systems, where $D\sim\operatorname{poly}(N)$, but for two- and three-dimensional systems, the computational cost scales exponentially as $\operatorname{exp}(\sqrt{N})$ and $\operatorname{exp}(N^{2/3})$, respectively.  

Anisotropic interactions in spin models can give rise to incommensurate phases, where the periodicity of magnetic moments does not align with the underlying crystal lattice in a simple rational ratio. This means that the magnetic ordering wave vector is not a simple fraction of the reciprocal lattice vectors of the crystal, often resulting in a zero net magnetic moment. In extended Kitaev models at zero temperature, two types of incommensurate phases can emerge: spiral order and quantum paramagnetic (QPM) order. The incommensurate spiral order (often simply referred to as "incommensurate") is a long-range ordered phase characterized by a nonzero local magnetic moment at each lattice site, while the total magnetization remains zero. This arises from continuously modulated spin arrangements, such as helical, cycloidal, or sinusoidal spin waves. It has been shown that the Kitaev honeycomb model with symmetric off-diagonal anisotropic $\Gamma$ nearest-neighbor (NN) interactions and symmetric diagonal isotropic $J$ NN interactions exhibits a rich phase diagram, including various incommensurate orders consistent with the observed ground states of known Kitaev materials~\cite{Winter2017,Rousochatzakis2024}.

Another incommensurate phase with respect to the crystal lattice is the QPM phase, which exhibits short-range order and a zero net magnetic moment at every lattice site. The QPM phase has been found in various Kitaev-based models, including the Kitaev-$\Gamma$ model in the absence of a magnetic field~\cite{Gohlke2018}, the ferromagnetic Kitaev model~\cite{Kaib2018}, the Kitaev-$\Gamma$-$\Gamma'$ model~\cite{Gohlke2020}, and the Kitaev-$\Gamma$-$J_1$-$J_3$ model in the presence of a magnetic field~\cite{Winter2018}. Our DMRG study reveals that anisotropic Ising interactions can induce QPM phases in the Kitaev honeycomb model even in the absence of a magnetic field.

In this work, we use the DMRG method to study the Kitaev-Ising-$J_1$-$J_2$ model, which maps the Kitaev honeycomb model onto a square lattice of superconducting qubits. This mapping introduces additional XX Ising-like interactions between NN and NNN sites due to parasitic coupling between qubits. We have calculated the ground state of the model using DMRG procedure for the $10\times10$ lattice with the periodic boundary conditions along the $x$-axis and open boundary condition along the $y$-axis. The phase diagram consists of two QPM phases, in addition to KSL and magnetic commensurate phases, such as ferromagnetic (FM), Neel, x- and y-stripy orders. We analyze the scaling behavior of entanglement entropy and the maximum bond dimension of MPS across all phases. Additionally, we propose a new optimization algorithm for DMRG to stabilize the ground state in low-entanglement phases.

This paper is organized as follows. In Sec.~\ref{sec:m_m}, we introduce the model and give a brief review of DMRG. In Sec.~\ref{sec:num_res}, we present the numerical results for the Kitaev-Ising-$J_1$-$J_2$ model. Specifically, we analyze the quantum phase diagram and spin-structure factor (Sec.~\ref{sec:phase_diagram}), examine the entanglement entropy and the maximum bond dimension across different phases (Sec.~\ref{sec:scaling_law}), and introduce the optimization algorithm (Sec.~\ref{sec:optimization}). And finally, we summarize our findings in Sec.~\ref{sec:dis}. 

\section{Model and method}
\label{sec:m_m}

In this section, we introduce the extended Kitaev model with NN and NNN XX coupling, and provide a brief overview of the DMRG method.

\subsection{Model}
\label{sec:m_m_model}

\begin{figure}[h!]
    \centering
    \includegraphics[width=0.7\linewidth]{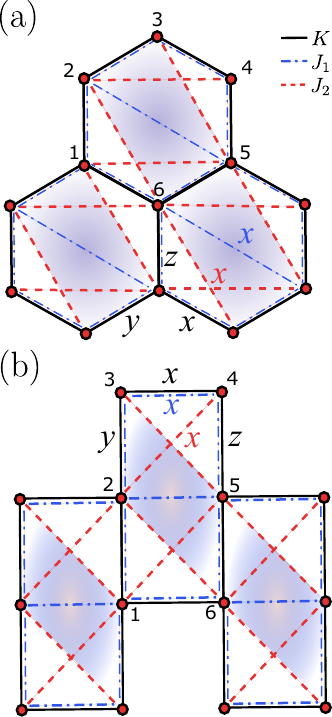}
    \caption{The Kitaev-Ising-$J_1$-$J_2$ model on the hexagonal (a) and square (b) lattices. Black lines represent the pure Kitaev model coupling $K$, blue dashed lines with dots indicate to the NN XX coupling $J_1$, and red dashed lines denote to the NNN XX coupling $J_2$. Shaded sectors denote the sites involved in the calculation of magnetic order parameters~\eqref{eq:sigma_order_1},\eqref{eq:sigma_order_2}.}
    \label{fig:lattice}
\end{figure}

The Kitaev honeycomb model~\cite{Kitaev2006} is described by the Hamiltonian 
\begin{equation}\label{eq:k_ham}
    H_\text{Kitaev} = J_x\!\!\sum_{\left<ij\right>_x}\!\!S^{x}_iS^{x}_j+J_y\!\!\sum_{\left<ij\right>_y}\!\!S^y_iS^y_j+J_z\!\!\sum_{\left<ij\right>_z}\!\!S^z_iS^z_j,
\end{equation}
where $i$, $j$ label the sites on a hexagonal lattice, and $\left<ij\right>_{x,y,z}$ denotes the NN bond along the $x-$,$y-$, and $z-$type links, respectively. The model has no continuous global spin symmetry, but it has a rich local symmetry. A specific product of spin operators associated with an elementary hexagon  $W_\text{p}=2^6S^x_1S^y_2S^z_3S^x_4S^y_5S^z_6$, commutes with the full Hamiltonian and is referred to as the plaquette operator (see Fig.~\ref{fig:lattice}). This model is exactly solvable, and the ground state corresponds to the 'flux-free' regime with $W_\text{p}=+1$ for all plaquettes. Assuming $J_x=J_y=J_z=K$, the Hamiltonian~\eqref{eq:k_ham} simplifies to
\begin{equation}\label{eq:k_ham_k}
    H_\text{Kitaev}=K\sum_{\left<ij\right>_a}S^{\gamma}_iS^{\gamma}_j,
\end{equation}
where $\gamma=\left\{x,y,z\right\}$ and $a=\left\{x,y,z\right\}$. In this parameter regime, the ground state of the pure Kitaev model is a gapless Kitaev spin liquid (KSL).

In this work, we map the Kitaev model to the square lattice of superconducting qubits with the same coupling topology as in the Hamiltonian~\ref{eq:k_ham_k} (see Fig.~\ref{fig:lattice}), and due to parasitic interactions between qubits, we incorporate both NN and NNN Ising interactions, leading to the Hamiltonian 

\begin{equation}\label{eq:extra_ham}
    H_\text{extra}=J_1\!\!\sum_{\left<ij\right>}\!S^x_iS^x_j+J_2\!\!\!\sum_{\left<\left<ij\right>\right>}\!\!S^x_iS^x_j,
\end{equation}
where $\left<ij\right>$ and $\left<\left<ij\right>\right>$ stand for the first and the second NN bonds, $J_1$ and $J_2$ are the NN and NNN coupling constants. This defines the Kitaev-Ising-$J_1$-$J_2$ model, which extends the Kitaev honeycomb model by incorporating Ising-like XX interactions on NN and NNN bonds after mapping to a square lattice. The full Hamiltonian is

\begin{align}\label{eq:ham}
    H & \nonumber = H_\text{Kitaev}+H_\text{extra}&\\
      & = K\sum_{\left<ij\right>_a}S^{\gamma}_iS^{\gamma}_j + J_1\!\!\sum_{\left<ij\right>}\!S^x_iS^x_j+J_2\!\!\!\sum_{\left<\left<ij\right>\right>}\!\!S^x_iS^x_j,
\end{align}
where the coupling constants are parameterized as
\begin{equation}\label{eq:param}
    K=\cos{\varphi},\,\,J_1=\sin{\varphi}\cos{\alpha},\,\,J_2=\sin{\varphi}\sin{\alpha},
\end{equation}
where $\varphi$ changes from $0$ to $2\pi$ and $\alpha$ changes from $0$ to $\pi$. Note, that we do not consider $\alpha$ from $0$ to $2\pi$, because in the span from $\pi$ to $2\pi$ there is a nothing new, just repeating the same picture as for the $\alpha$ from $0$ to $\pi$.


We define the following order parameters to characterize magnetic phases
\begin{align}\label{eq:sigma_order_1}
    \sigma_\text{I}&=\frac{1}{4}\left(\left<S^x_2S^x_3\right>+\left<S^x_3S^x_5\right>+\left<S^x_5S^x_6\right>+\left<S^x_6S^x_2\right>\right),\\
    \sigma_\text{II}& {} = \frac{1}{4}\left(-\left<S^x_2S^x_3\right>+\left<S^x_3S^x_5\right>-\left<S^x_5S^x_6\right>+\left<S^x_6S^x_2\right>\right),\label{eq:sigma_order_2}
\end{align}
 where indices 1,..,6 correspond to sites on a hexagon in the lattice (see Fig.~\ref{fig:lattice}). In the presence of the topological regime (KSL) there is no magnetic order and $\sigma_\text{I}=\sigma_\text{II}=0$, but the specific spin product $W_p=2^6S^x_1S^y_2S^z_3S^x_4S^y_5S^z_6=1$ will be a topological order parameter (see Sec.~\ref{sec:phase_diagram}). 

\subsection{Density Matrix Renormalization Group}
\label{sec:m_m_method}

Here, we briefly summarize the key aspects of the DMRG method and the MPS ansatz. For a more comprehensive discussion, see~\cite{Schollwock2011,Stoudenmire2012,Catarina2023,Ganahl2023}.

Consider a spin-$\frac{1}{2}$ lattice with $N$ sites, with each site described by a basis of spin-up/spin-down states $\left\{\left|\uparrow\right>,\left|\downarrow\right>\right\}$. The many-body wave function of the system is given by
\begin{equation}\label{eq:wave_function}
    \left|\psi\right>=\sum_{i_1i_2...i_N}\psi^{i_1i_2..i_N}\left|i_1i_2...i_N\right>,
\end{equation}
where $\psi^{i_1i_2..i_N}$ represents $2^N$ complex amplitudes and $\left|i_1i_2...i_N\right>$ stands for the product basis $\left|i_1\right>\otimes\left|i_2\right>\otimes\dotsm\otimes\left|i_N\right>$ for $2^N$-dimensional vector space of the $N$ sites. Similarly, the local many-body Hamiltonian in this basis is expressed as
\begin{equation}\label{eq:generic_hamiltonian}
    \mathcal{H}=\sum_{\left\{i\right\},\left\{j\right\}}\mathcal{H}^{i_1j_1i_2j_2..i_Nj_N}\left|i_1i_2...i_N\right>\left<j_1j_2...j_N\right|.
\end{equation}

Our goal is to compute an accurate approximation of the ground state $\left|\psi_\text{GS}\right>$ of the Hamiltonian $\mathcal{H}$. One can use the exact diagonalization method, but we are limited to small systems, because the dimension of the Hilbert space~\eqref{eq:wave_function} grows exponentially with the system size $N$. This limitation is known as the exponential wall problem\cite{Kohn1999}. To overcome this challenge, we require a computational method that avoids the need to explicitly store the full wave function~\eqref{eq:wave_function} amplitudes.

One of the numerical methods that allows bypassing the exponential wall problem is DMRG~\cite{White1992,White1993}. This algorithm has its roots in Wilson's numerical renormalization group~\cite{Wilson1975,Bulla2008}, whose key idea is to divide a large quantum system into smaller blocks that can be solved by exact diagonalization. This algorithm is not very effective for strongly correlated systems~\cite{White1993}. 

In the DMRG method the truncation procedure is more accurate than in previous approaches, because the basis is rotated such that only a small number of states are required to represent the ground state, while the remaining states can be discarded. To improve efficiency, basis rotations are performed iteratively on a few sites at a time, treating the remaining sites as an environment. This iterative sweeping process enables global basis optimization across all lattice sites without requiring a full Hilbert space transformation in a single step. As a result, the variational algorithm produces a wave function expressed in a compact form known as the matrix product state (MPS)~\cite{Fannes1992}. The modern formulation of DMRG is built upon tensor networks, where the algorithm begins with an MPS as a variational ansatz and then optimizes all its coefficients~\cite{Ostlund1995,Dukelsky1998,Schollwock2011}.

The computational cost of optimizing MPS from site $n$ to site $n+1$ is $O(D^3)$, resulting in a total optimization sweep cost that scales as $O(ND^3)$, where $D$ is the bond dimension and $N$ is the system size. For a generic (nonlocal) Hamiltonian~\eqref{eq:generic_hamiltonian} bond dimension $D$ grows exponentially with the entanglement entropy, following $D\sim\exp{\left(S\right)}$~\cite{Vidal2003,Peschel2003}. Simultaneously, the entanglement entropy obeys a scaling law known as the volume law, where $S\sim N$. Consequently, in the generic case, the bond dimension scales exponentially with system size as $D\sim\exp{\left(N\right)}$. 

A key factor in the success of DMRG is the area law: the ground state of many physical systems can often be accurately represented by an MPS with a fixed bond dimension $D$~\cite{Vidal2003,Srednicki1993,Hastings2007,Amico2008,Eisert2010}. This implies that the entanglement entropy $S$, rather than being an extensive quantity, is at most proportional to the boundary of the bipartition. For the ground state of a local Hamiltonian~\eqref{eq:generic_hamiltonian} describing a $d$-dimensional system of linear size $L$ made of $N=L^d$ sites, the entanglement entropy scales as 
\begin{equation}
    S\sim L^{d-1}=N^{\left(d-1\right)/d},
\end{equation}
\begin{equation}
    D\sim\exp{\left(L^{d-1}\right)}=\exp{\left(N^{\left(d-1\right)/d}\right)}.
\end{equation}

In one dimension ($d=1$), the entanglement entropy remains constant $S\sim\operatorname{const}$ for gapped systems, implying that the bond dimension $D$ is also constant and independent of system size $N$. Consequently, the computational cost for a $1D$ scales as $O(N)$. For 1D gapless/critical models, the entanglement entropy follows a logarithmic scaling with system size $S\sim\operatorname{ln}{(N)}$, leading to a polynomially growing bond dimension $D\sim\operatorname{poly}{(N)}$. This behavior is known as the logarithmic correction to the area law~\cite{Vidal2003}.

In higher dimensions ($d=2,3$) the computational complexity increases significantly. While the system still obeys the area law, the bond dimension grows exponentially with the system size as $\exp{(\sqrt{N})}$ and $\exp{(N^{2/3})}$ for $d=2,3$. Nevertheless, DMRG remains advantageous, as its computational cost scales as a fractional power of $N$, in contrast to the $\exp{(N)}$ scaling of brute-force computation methods, which suffer from the exponential wall problem~\cite{Kohn1999}. 

In conclusion, DMRG is a highly effective computational method that has become dominant in the study of strongly correlated 1D systems.  Despite its unfavorable scaling in higher dimensions, DMRG remains widely used for 2D systems, particularly when one dimension is relatively small, as in stripe and cylindrical geometries~\cite{White2000,Weng2006,Vidal2013,Jiang2019} (for a review of 2D DMRG, see~\cite{Stoudenmire2012}). Additionally, DMRG has been successfully applied to small 3D molecular systems~\cite{Chan2002,Legeza2003,Chan2004,Moritz2004,Goings2022}.

\section{Numerical results}
\label{sec:num_res}

In this section, we present the numerical results for the Kitaev-Ising-$J_1$-$J_2$ model~\eqref{eq:ham} obtained using DMRG simulations. In particular, we determine the ground-state phase diagram of the system on a $10\times10$ square lattice with $N=100$ sites and with the periodic boundary conditions along $x$-axis and open boundary condition along $y$-axis (Sec.~\ref{sec:phase_diagram}). We compute the entanglement entropy and the maximum bond dimension as functions of system size (Sec.\ref{sec:scaling_law}). Additionally, we introduce an analog of adiabatic quantum optimization for DMRG to prevent getting stuck in the local minima and reach a global energy minimum (Sec.~\ref{sec:optimization}). DMRG simulations were performed with a truncation error less than $10^{-9}$, the Krylov dimension is 5, and the number of sweeps is 35. All calculations were carried out using our proposed optimization algorithm (see Sec.~\ref{sec:optimization}).

\subsection{Quantum phase diagram}
\label{sec:phase_diagram}

We calculate the order parameters $\sigma_\text{I}$~\eqref{eq:sigma_order_1}, $\sigma_\text{II}$~\eqref{eq:sigma_order_2}, and topological order parameter $W_p=2^6S^x_1S^y_2S^z_3S^x_4S^y_5S^z_6$, as introduced in Sec.~\ref{sec:m_m_model} and summarized in Tab.~\ref{tab:order}. The order parameters $\sigma_\text{I}$ and $\sigma_\text{II}$ indicate the presence of a magnetic order, whereas $W_p$ characterizes the topological order. As shown in Fig.~\ref{fig:diagram}, the ground state hosts seven distinct phases: four with commensurate order (FM, Neel, x- and y-stripy), two with quantum paramagnetic order (labelled QPM1 and QPM2), and KSL phase. We calculate the static spin-structure factor
\begin{equation}\label{eq:ss_factor}
    S(\mathbf{k})=\frac{1}{N}\sum_{ij}\left(\left<S^x_iS^x_j\right>+\left<S^y_iS^y_j\right>+\left<S^z_iS^z_j\right>\right)e^{i\mathbf{k}\left(\mathbf{r}_i-\mathbf{r}_j\right)},
\end{equation}
where $N$ is the number of sites, and $\mathbf{r}_i$ denotes the position of the site $i$.     

Spin-structure factor characterizes the spin-spin correlations in the system. In Fig.~\ref{fig:ss_factor} the Bragg peaks corresponding to the $\left<\mathbf{S}_i\mathbf{S}_j\right>$ correlations are plotted in the extended Brillouin zone. For commensurate (ordered) magnetic phases, the Bragg peak positions reflect the underlying lattice symmetry. In contrast, in the QPM phases, which are incommensurate, the Bragg peaks are shifted and do not follow the lattice symmetry. Thus, the magnetic structure can be inferred from the Bragg peak positions in reciprocal space. Additionally, we note that the spin-structure factor can provide insight into the system's entanglement properties~\cite{Cramer2011}, which we discuss in Sec.~\ref{sec:dis}.

We compute the spin-spin correlator $\left<\mathbf{S}_i\mathbf{S}_j\right>$ between sites $i$ and $j$. This value identifies the decay of the spin correlations with a distance $\left|i-j\right|$ and is shown in Fig.~\ref{fig:ss_corr}. In commensurate magnetic phases, the spin-spin correlation function remains nearly constant with distance, indicating long-range order. In contrast, KSL and QPM phases exhibit short-range order. Specifically, the QPM phase is characterized by an exponential decay of correlations, while in the KSL phase, spin correlations vanish beyond the second-neighbor distance~\cite{Baskaran2007}.  

It is important to clarify how phase boundaries are determined in Fig.~\ref{fig:diagram}. In general, there are three possible types of phase transitions.

The first type occurs between the topological KSL phase and a topologically trivial phase (any other phase). In this case, the topological order parameter $W_p$ drops sharply to zero at the transition point. Therefore, we define the phase transition as occurring where $W_p=0.5$. A similar scenario applies to transitions between different commensurate magnetic phases, where the magnetic order parameter $\sigma_\text{I,II}$ takes descrite values ($0$ or $\pm0.5$) at the transition point, depending on the phase (see Tab.~\ref{tab:order}). Thus, in both cases, we assume the phase transition occurs when the order parameter decreases by half.

A different behavior is observed in transitions from the QPM phase to a commensurate magnetic phase. Unlike the previous cases, there is no sharp discontinuity in the order parameter, making this transition resemble a crossover rather than a true phase transition. By analyzing the spatial distribution of the average magnetic moment at each lattice site, we identify the transition as occurring when $\left|\sigma_\text{I, II}\right|<0.4$. Note that the QPM-KSL transition follows the first case described above. 
 
\subsubsection{Kitaev Spin Liquid state}

As mentioned in Sec.~\ref{sec:m_m_model}, our model includes the Kitaev term~\eqref{eq:k_ham}, which gives rise to either a gapped or gapless KSL state. For simplicity, we assume that $J_x=J_y=J_z=K$ in the Kitaev term~\eqref{eq:k_ham}, corresponding to the gapless KSL phase. In this phase, the magnetic order parameters~\eqref{eq:sigma_order_1} and~\eqref{eq:sigma_order_2} vanish (see Tab.~\ref{tab:order}), but there is a topological order parameter $W_p=2^6S^x_1S^y_2S^z_3S^x_4S^y_5S^z_6$ (see Fig.~\ref{fig:lattice}) remains nontrivial.  In the pure Kitaev model, this parameter takes the value $W_p=1$. However, in extended Kitaev models, $W_p=1$ cannot be maintained throughout the entire KSL phase and exhibits fluctuations~\cite{Liu2020}.

The KSL state exists within a narrow range of parameters near the Kitaev points $\varphi=\{0,\pi,2\pi\}$. This result is expected, as a similar phase structure is typical in spin models with Kitaev interactions, such as the Kitaev-Heisenberg model~\cite{Reuther2011,Schaffer2012}, the $K_1$-$K_2$ model~\cite{Rousochatzakis2015}, and the Kitaev-Heisenberg-Gamma model (or J-K-$\Gamma$ model)~\cite{Chaloupka2013,Rau2014}.

As shown in Fig.~\ref{fig:ss_factor}, two distinct spin liquid phases emerge near the Kitaev points, depending on the sign of the Kitaev interaction coupling constant $K=\cos{\varphi}$. For $K>0$, the system enters the ferromagnetic Kitaev spin liquid (FM KSL) phase (see Fig.~\ref{fig:ss_factor}a), characterized by a soft peak at the $\Gamma$-point. Conversely, for $K<0$, the system realizes the antiferromagnetic Kitaev spin liquid (AFM KSL) phase (see Fig.~\ref{fig:ss_factor}b), where the spin-structure factor exhibits broad maxima around the K-points of the first Brillouin zone (BZ), analogous to the FM KSL phase. In both regimes, the KSL phase is marked by delocalized Bragg peaks in the spin-spin correlations (see Fig.~\ref{fig:ss_factor}(a-b)), which may indicate strong quantum entanglement (see Sec.~\ref{sec:dis}). Throughout this work, we refer to both phases as KSL, as they share the same fundamental properties. Notably, as a quantum spin liquid (QSL) phase, KSL exhibits only short-range order. As shown in Fig.~\ref{fig:ss_corr}, the spin-spin correlation function remains nonzero only for nearest neighbors~\cite{Baskaran2007}.

\subsubsection{Commensurate magnetic phases}

Well-known phases in frustrated magnetic models are commensurate magnetic phases such as FM, Neel, $x$-stripy, and $y$-stripy (see Fig.~\ref{fig:diagram} and Fig.~\ref{fig:ss_factor}). In this work, we refer to the phases as $x$-stripy and $y$-stripy in the context of the honeycomb model mapped to the square lattice, as this notation aligns more naturally with the spin configuration shown in real space in Fig.~\ref{fig:diagram}. However, for the honeycomb lattice, the $y$-stripy phase is more commonly referred to as the zigzag phase.

As summarized in Tab.~\ref{tab:order}, the topological order parameter vanishes in all magnetic phases, while each magnetic phase exhibits a distinct value of the order parameters~\eqref{eq:sigma_order_1} or~\eqref{eq:sigma_order_2}. The spin-spin correlation function $\left<\mathbf{S}_i\mathbf{S}_j\right>$ remains unchanged for next-nearest neighbors, even at large distances, indicating long-range order (see Fig.~\ref{fig:ss_corr}). The Bragg peaks of the spin-spin correlations are localized and preserve the honeycomb lattice symmetry in the reciprocal space (see Fig.~\ref{fig:ss_factor}(e-h)). This localization may indicate low entanglement in these phases (see Sec.~\ref{sec:dis}).

\subsubsection{Quantum paramagnetic phases}

In general, strong frustration in extended Kitaev models gives rise to phases that are incommensurate with the lattice symmetry. These phases can manifest as either incommensurate spiral magnetic order or a quantum paramagnetic (QPM) phase. Unlike the spiral order, the QPM phase exhibits only short-range order and has a zero average magnetic moment at every lattice site. Previous studies have identified the QPM phase in Kitaev models with symmetric off-diagonal anisotropic $\Gamma$ and $\Gamma'$ interactions~\cite{Gohlke2018} in the absence of a magnetic field. In contrast, our model includes only diagonal anisotropic interactions (Kitaev and Ising $J_1$, $J_2$) and lacks any $\Gamma$-like interactions. This results in the emergence of the QPM phase without an accompanying incommensurate spiral order in the phase diagram.

\begin{figure}[h!]
    \centering
    \includegraphics[width=\linewidth]{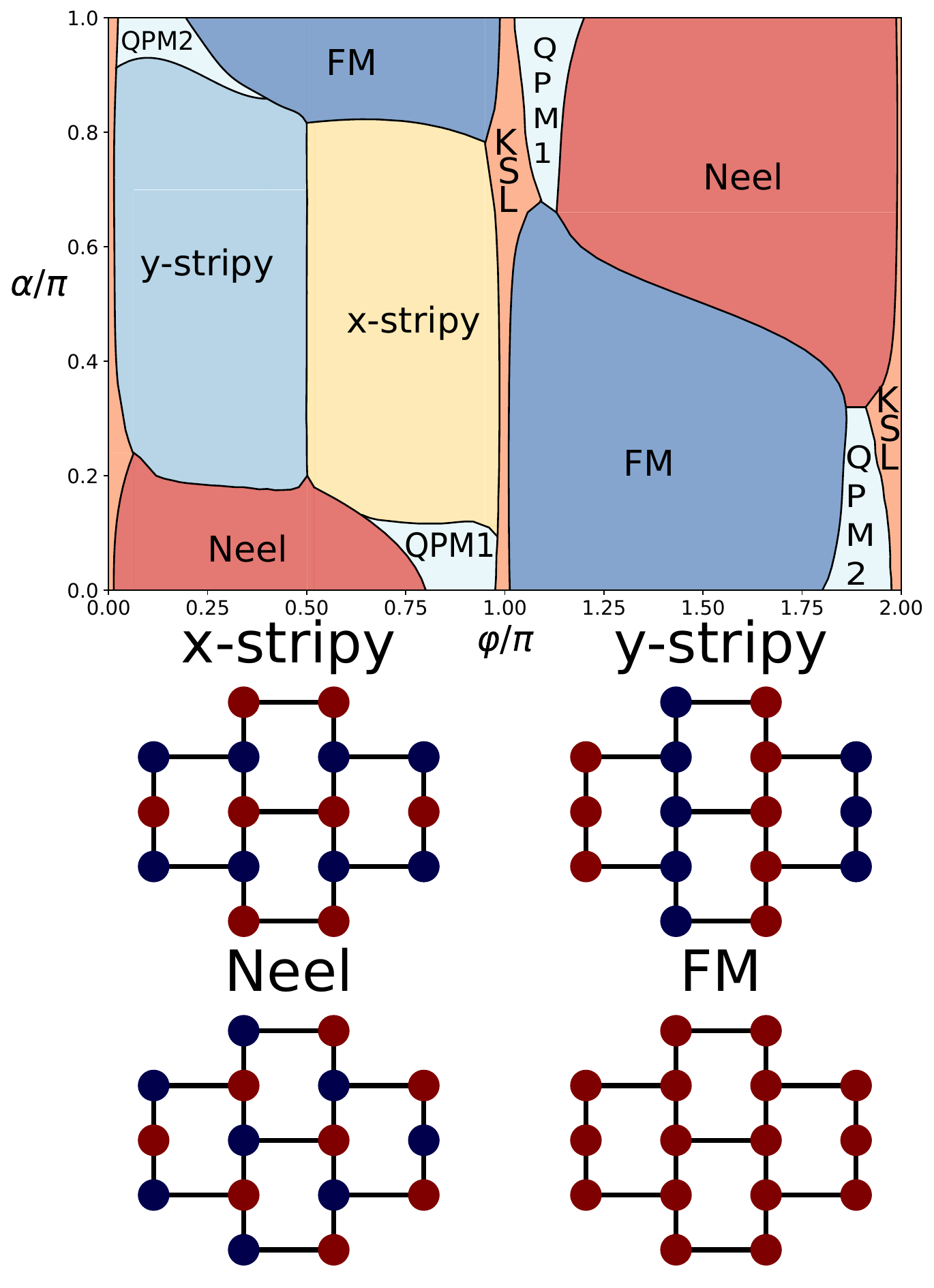}
    \caption{Phase diagram of the Kitaev-Ising-$J_1$-$J_2$ model, along with schematic spin configurations for commensurate phases. Red (blue) points represent spin-up (spin-down) states with $\left<S^x\right>=+1/2$ ($\left<S^x\right>=-1/2$).}
    \label{fig:diagram}
\end{figure}

\begin{figure}[h!]
    \centering
    \includegraphics[width=\linewidth]{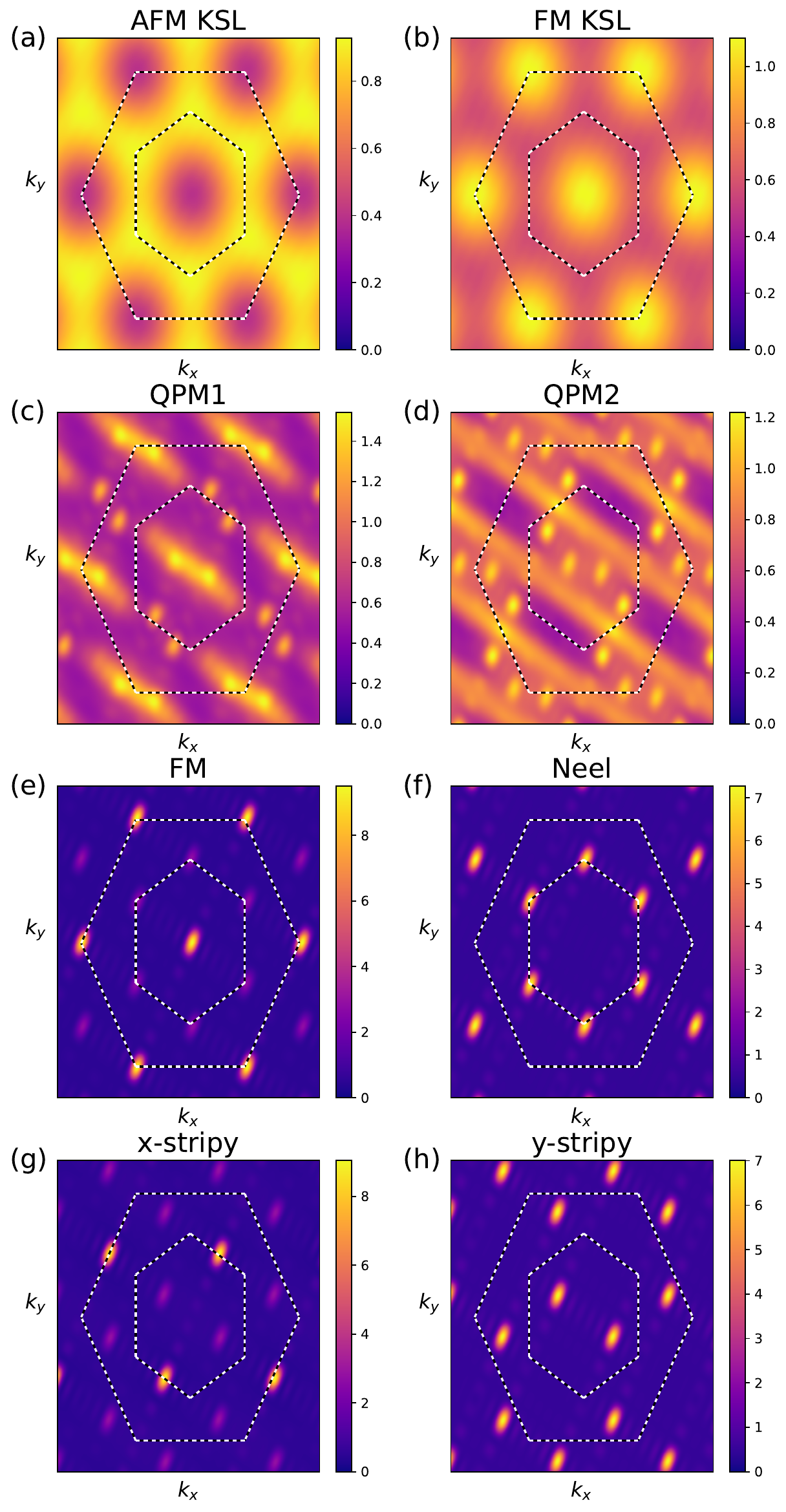}
    \caption{Spin-structure factor $S(\mathbf{k})$ of the Kitaev-Ising-$J_1$-$J_2$ model for representative momenta in different phases. The larger black and white dashed hexagone denotes the extended Brillouin zone, while the inner hexagone represents the first Brilloin zone. (a) AFM KSL phase, plotted at $\alpha=0,\,\,\varphi=0$. (b) FM KSL phase at $\alpha=0,\,\,\varphi=\pi$. (c) QPM1 phase at $\alpha=0,\,\,\varphi=0.85\pi$. (d) QPM2 phase at $\alpha=0,\,\,\varphi=1.85\pi$. (e) FM phase at $\alpha=0,\,\,\varphi=1.5\pi$. (f) Neel phase $\alpha=0,\,\,\varphi=0.5\pi$. (g) $x$-stripy phase at $\alpha=0.5\pi,\,\,\varphi=0.75\pi$. (h) $y$-stripy phase at $\alpha=0.5\pi,\,\,\varphi=0.25\pi$. }
    \label{fig:ss_factor}
\end{figure}

\begin{figure}[h!]
    \centering
    \includegraphics[width=\linewidth]{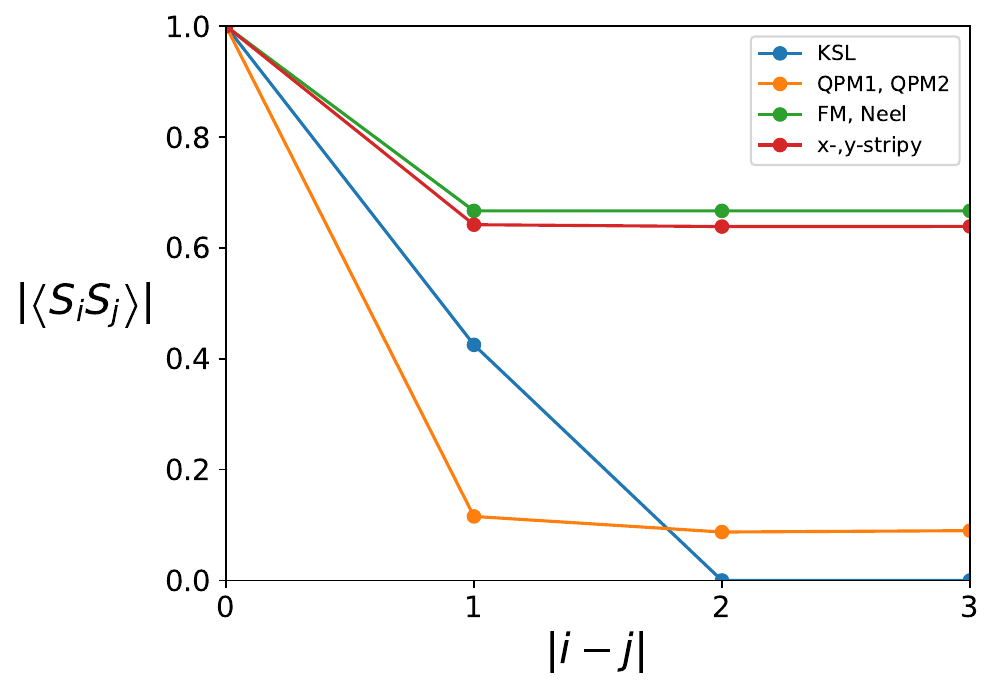}
    \caption{The spin-spin correlation function is plotted as a function of the distance $\left|i-j\right|$ between spins. The calculations for different phases were performed at the same values of $\alpha$ and $\varphi$ as in the Fig.~\ref{fig:ss_factor}.}
    \label{fig:ss_corr}
\end{figure}

The QPM state lacks topological order ($W_p=0$) and have spatially modulated order parameters (see Tab.~\ref{tab:order}). These phases stabilize near the Kitaev limits and exhibit a zero average magnetic moment at each lattice site. The first QPM phase (QPM1, see Fig.~\ref{fig:diagram} and Fig.~\ref{fig:ss_factor}c) is incommensurate along the $\mathbf{b_1}=\frac{2\pi}{3}(\sqrt{3},1)$ direction. The second QPM phase (QPM2, see Fig.~\ref{fig:diagram} and Fig.~\ref{fig:ss_factor}d) is incommensurate along both reciprocal lattice basis vectors $\mathbf{b_1}$ and $\mathbf{b_2}=\frac{4\pi}{3}(1,0)$. Both QPM phases are highly degenerate and frustrated, which leads to strong entanglement comparable to the entanglement in the KSL state (see Fig.~\ref{fig:entr_and_bond_dim})~\cite{Sorensen2021}. The Bragg peaks of the spin-spin correlations are partially delocalized and shifted (see Fig.~\ref{fig:ss_factor}(c-d)), indicating incommensurate magnetic order. Together with short-range order, characterized by the exponential decay of the spin-spin correlation function with distance (see Fig.\ref{fig:ss_corr}), and a zero average magnetic moment at each lattice site, these features are hallmarks of the QPM phase. Note that this partial delocalization may serve as a signature of a highly entangled state (see Sec.~\ref{sec:dis}).

\begin{table}[h!]
    \caption{Order parameters and entanglement entropy for all possible phases in the Kitaev-Ising-$J_1$-$J_2$ model.}
    \label{tab:order}
    \centering
    \begin{ruledtabular}
    \begin{tabular}{c c c c c}
            \textrm{Phase} & $W_p$ & $\sigma_\text{I}$ & $\sigma_\text{II}$ & $S$\\
            \colrule
            KSL & 1 & 0 & 0 & $\sim\sqrt{N}$\\
            QPM1 & 0 & $-0.1\div-0.4$ & $0.1\div0.4$ & $\sim\sqrt{N}$\\
            QPM2 & 0 & $0.1\div0.4$ & $-0.1\div-0.4$ & $\sim\sqrt{N}$\\
            FM  & 0 & 1 & 0 & $\sim\operatorname{const}$\\
            $x$-stripy & 0 & -1 & 0 & $\sim\operatorname{const}$\\
            Neel & 0 & 0 & 1 & $\sim\operatorname{const}$\\
            $y$-stripy & 0 & 0 & -1 & $\sim\operatorname{const}$\\ 
    \end{tabular}
    \end{ruledtabular}

\end{table}

\subsection{Entanglement entropy and bond dimension}
\label{sec:scaling_law}

\begin{figure}[h!]
    \centering
    \includegraphics[width=\linewidth]{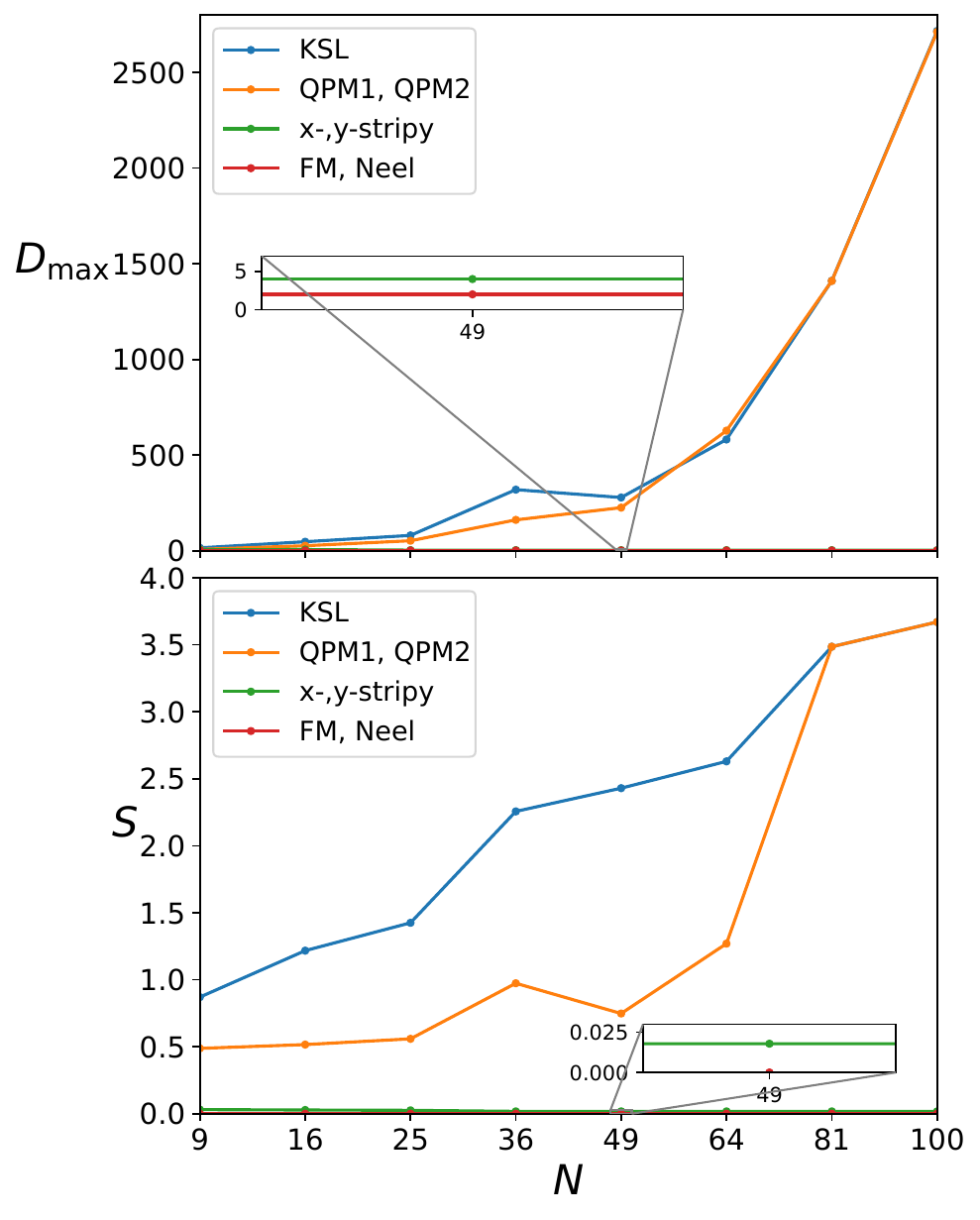}
    \caption{Scaling of the entanglement entropy (top) and the maximum bond dimension (bottom) with system size. The calculations for different phases were performed at the same values of $\alpha$ and $\varphi$ as in the Fig.~\ref{fig:ss_factor}. The lattice size $N$ varies from $3\times3$ to $10\times10$.}
    \label{fig:entr_and_bond_dim}
\end{figure}

 In this section, we analyze the entanglement entropy and the maximum bond dimension across different phases. These quantities are plotted in Fig.~\ref{fig:entr_and_bond_dim} as functions of system size.

The most entangled phases are the KSL and the QPM phases QPM1 and QPM2 (see Tab.~\ref{tab:order} and Fig.~\ref{fig:entr_and_bond_dim}). In these phases, the entanglement entropy $S$ scales as $\sqrt{N}$, leading to an exponentially growing bond dimension $D\sim e^{\sqrt{N}}$, in accordance with the area law. Interestingly, incommensurate phases exhibit the same exponential scaling, a behavior typically associated with the KSL state~\cite{Zhang2011,Meichanetzidis2016}. Since the MPS representation grows exponentially with system size, more advanced wave function ansätze, such as projected entangled pair states (PEPS)~\cite{Verstraete2004,Verstraete2006} and the multiscale entanglement renormalization ansatz (MERA)~\cite{Vidal2007,Cincio2008}, are required for efficient simulations in 2D (about PEPS and MERA see nice reviews~\cite{Stoudenmire2012,Cirac2021}). Notably, despite its exponential scaling, this approach remains significantly more efficient than brute-force methods, which scale as $e^{N}$. 

Commensurate magnetic phases have a near-zero entanglement entropy value,see Fig.~\ref{fig:entr_and_bond_dim}, resulting in a constant bond dimension $D$. In practice, this implies that these low-entanglement magnetic phases are well described by mean-field theory, requiring only a small bond dimension to accurately capture the MPS representation. After the first sweep, the MPS is already well-converged. However, this scenario presents a challenge: in our system, the DMRG algorithm tends to become trapped in local minima.

\subsection{Optimization algorithm for ground state energy}
\label{sec:optimization}

\begin{figure}
    \centering
    \includegraphics[width=\linewidth]{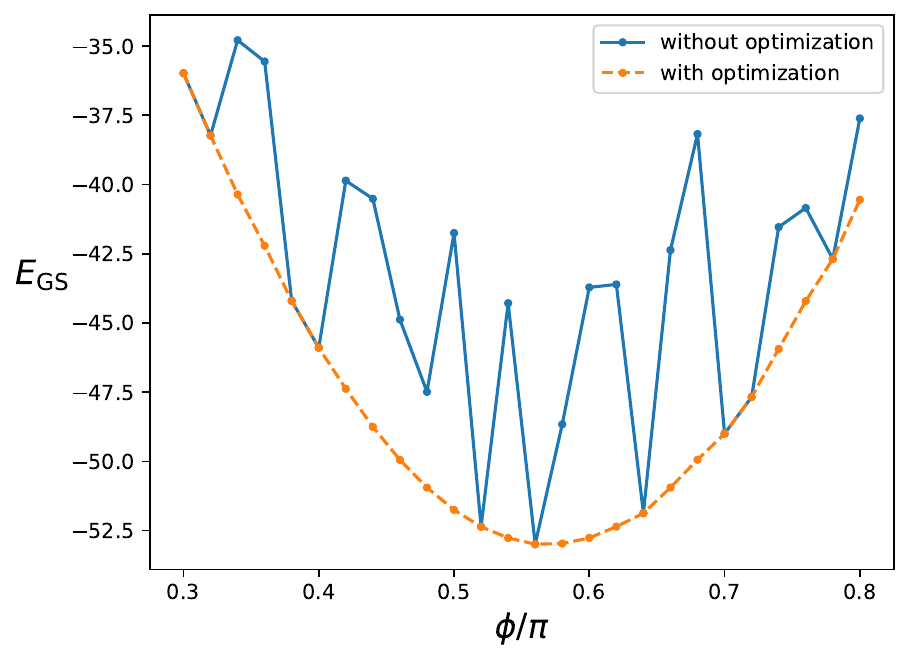}
    \caption{Results of the optimization algorithm for $\varphi/\pi\in\left[0.3,0.8\right]$ and $\alpha=0.24\pi$. The optimization procedure is effective across the entire parameter range of $\varphi$ and $\alpha$. The blue line represents the case without optimization, while the dashed orange line shows the ground-state energy obtained with optimization.}
    \label{fig:optimization}
\end{figure}

In low-entanglement phases, the DMRG method encounters a convergence issue —- the ground state energy fails to stabilize (see the blue line in Fig.~\ref{fig:optimization}). Potential solutions, such as introducing a noise term, increasing the dimension of the Krylov subspace, or extending the number of Krylov subspace rebuilds, either prove ineffective or further exacerbate the problem.

Here, we propose an algorithm to efficiently determine the ground state energy. We introduce the following modified Hamiltonian
\begin{equation}\label{eq:optim_ham}
    H(\lambda)=\left(1-\lambda\right)H_0+\lambda H', 
\end{equation}
where $\lambda$ is the weight parameter, the $H_0$ is the origin Hamiltonian~\eqref{eq:ham}, and the $\lambda H'$ represents a perturbation term. We consider the following perturbation term
\begin{equation}\label{eq:perturb_term}
    H'=\sum_i\mathbf{B}\cdot\mathbf{S}_i,
\end{equation}
where $B$ is a magnitude of an applied magnetic field. In our algorithm, we set $B=(0,0,B_\text{z})$. Notably, the optimization procedure remains effective for any choice of magnetic field direction in the perturbation term~\eqref{eq:perturb_term}.

Now we are ready to perform the optimization algorithm, which has the following structure. In the first step, we set $\lambda=1$ in the Hamiltonian~\eqref{eq:optim_ham} and perform a DMRG routine to obtain the optimized wave function $\left|\psi_{\lambda=1}\right>$. We use  $\left|\psi_{\lambda=1}\right>$ as an initial in the next step, where we gradually (adiabatically) "turn off" the perturbation term by reducing the weight parameter $\lambda$ by a small $\varepsilon$ and after DMRG procedure we get a new optimized wave function $\left|\psi_{\lambda=1-\varepsilon}\right>$. Then we repeat this process until $\lambda=0$. In the final step, we perform the DMRG for original Hamiltonian~\eqref{eq:ham}, using $\left|\psi_{\lambda=0}\right>$ as the initial wave function, to obtain the optimized ground state. 

In Fig.~\ref{fig:optimization} is shown the result of performing optimization algorithm, where the ground state energy is plotted with the respect to the parameter $\varphi$. As was written before, in the low entangled phase MPS is getting stuck. This occurs because commensurate magnetic phases (low entanglement states) are well described by a mean-field ansatz, resulting in a fixed bond dimension for each DMRG sweep. Consequently, the bond dimension cannot gradually increase, as the MPS becomes stuck in a low-energy state without reaching the true ground state. Our optimization algorithm addresses this issue by preparing the MPS in a controlled manner, ensuring that it converges directly to the ground state.

\section{Discussion}
\label{sec:dis}

In this article, we investigate the Kitaev honeycomb model with the additional XX Ising interaction between the nearest and the next nearest neighbors (Kitaev-Ising-$J_1$-$J_2$ model). Using DMRG, we determine the ground state of the system, which hosts low entangled commensurate magnetic phases and entangled KSL and unconventional QPM phases. We show that anisotropic Ising interactions can stabilize QPM phases, previously thought to require more complex interactions in the absence of the magnetic field. We analyze the scaling behavior of entanglement entropy and the bond dimension of the matrix product state, and we propose an optimization algorithm to improve DMRG convergence in low entangled phases. 

We note that the main motivation for the considered model is to simulate the Kitaev honeycomb model on a superconducting quantum processor. The goal of such simulations is to capture the KSL phase. We show that even a weak parasitic coupling leads to an instability of the KSL phase, which must be taken into account in real experiments and simulations.

As we mentioned in the Sec.~\ref{sec:phase_diagram}, high entangled phases (such as KSL, QPM1, and QPM2) have a delocalized spin-structure factor. This indicates not only that the spin-spin correlations in the system are incommensurate with the lattice, but is also an measure of the entanglement of the system. In the Ref.~\cite{Cramer2011}, is considered the quantity $E(\mathbf{q})=\operatorname{max}\{0,1-\frac{1}{2}S(\mathbf{q})\}$, which provides a lower bound to the many-particle entanglement (as measured in terms of the best separable approximation) contained in the system, and $\operatorname{min} \{E(\mathbf{q})\}$ shows the minimum possible entanglement in the system. We found that $\operatorname{min}\{E(\mathbf{q})\}=0$ in the low entangled phases (FM, Neel, $x$-, and $y$-stripy) and $\operatorname{min}\{E(\mathbf{q})\}\ne0$ in the high entangled phases (KSL, QPM1, and QPM2). This correlates with our results and suggests that delocalization of the spin-structure factor leads to the appearance of entanglement in the system.

\section*{Acknowledgments}
A.V.K thanks the support from the Foundation for the Advancement of Theoretical Physics and Mathematics “BASIS” and from the Russian Science Foundation under Grant No 22-72-10074. The DMRG calculations were performed using the ITensor Library~\cite{Fishman2022}.

\bibliography{ref}

\end{document}